\documentclass[prl,twocolumn,nofootinbib]{revtex4}
\usepackage{graphicx}
\usepackage{amssymb,amsmath,enumerate}
\usepackage{color}
\usepackage{bm}

\newcommand{\be}{\begin{eqnarray}}
\newcommand{\ee}{\end{eqnarray}}

\newcommand{\abs[1]}{\left\lvert #1 \right\rvert}

\begin{document}

\title{Computer simulation of random loose packings of micro-particles in presence of adhesion and friction}

\author{Wenwei Liu, Shuiqing Li \footnote{lishuiqing@tsinghua.edu.cn}, Sheng Chen}

\affiliation{ Key Laboratory for Thermal Science and Power
  Engineering of Ministry of Education, Department of Thermal
  Engineering, Tsinghua University, Beijing 100084, China }

\begin{abstract}

With a novel 3D discrete-element method specially developed with adhesive contact mechanics, random loose packings of uniform spherical micron-sized particles are fully investigated. The results show that large velocity, large size or weak adhesion can produce a relatively dense packing when other parameters are fixed, and these combined effects can be characterized by a dimensionless adhesion parameter ( $Ad=\omega/2\rho_pU^2_0R$). Four regimes are identified based on the value of $Ad$: RCP regime with $Ad<\sim 0.01$; RLP regime with $\sim 0.01<Ad<1$; adhesion regime with $1<Ad<20$ and an asymptotic regime with $Ad>20$. Force distribution of these adhesive loose packings follows $P(f)\sim f^\theta$ for small forces and $P(f)\sim \exp^{-\beta f}$ for big forces, respectively, which shares a similar form with that in packings without adhesion but results in distinct exponents of $\theta=0.879$, $\beta=0.839$. A local mechanical equilibrium analysis shows that adhesion enhances both sliding and rolling resistance so that fewer neighbours are needed to satisfy the force and torque balance.

\end{abstract}

\date{\today}

\maketitle

\section{1 Introduction}

Jammed particle packings have been studied to understand the microstructure and bulk properties of liquids, glasses and crystals \cite{Bernal59, Parisi10} and frictional granular materials \cite{Coniglio04, Andreotti13}. Two packing limits have been identified for disordered uniform spheres: the random close packing (RCP) and random loose packing (RLP) limits \cite{Bernal59, Scott60, Onoda90, Ciamarra08, Jerkins08, Dong06, Song08, Farrell10}. The upper RCP limit is reproduced for frictionless spheres at volume fractions $\phi\approx 0.64$ and has been associated with a freezing point of a first order phase transition \cite{Anikeenko07, Klumov11, Jin10, Panaitescu12}, among other interpretations \cite{Parisi10, Torquato10}. Only in the presence of friction, packings reach lower volume fraction up to the RLP limit $\phi_{RLP}\approx 0.55$ for mechanically stable packings \cite{Onoda90, Jerkins08, Farrell10}. However, most packings of dry small micrometer-sized particles in nature are not only subject to friction, but also {\it adhesion} forces as well. For instance, van der Waals forces generally dominate interactions between particles with diameters of around $10μm$ or smaller. In this case, the adhesive forces begin to overcome the gravitational and elastic contact forces acting on the particles and change macroscopic structural properties \cite{Li11, Marshall14}.

Despite the ubiquity of adhesive particle packings in almost all areas of engineering, biology, agriculture and physical sciences \cite{Marshall14, Dominik97, Blum04, Kinch07}, there have been few systematically investigations of these packings \cite{Yang:2000aa, Valverde04, Blum06, Martin:2008aa, Lois:2008aa, Parteli:2014aa}. The multi-coupling of adhesion, elastic contact forces and friction in the short-range particle-particle interaction zone and their further couplings with fluid forces (e.g. buoyancy, drag and lubrication) across long-range scales make it highly difficult to single out the effect of the adhesion forces, let alone to investigate the packing properties experimentally. With the progress of computer simulation techniques, discrete element simulation has become an efficient and accurate method to study the packing problems of micron-sized particles \cite{Yang:2000aa, Li11, Parteli:2014aa}, which are rather difficult to achieve in the experiment condition. In a dynamic packing process, particles will be arrested and eventually form a packing when all the kinetic energy is dissipated after a series of collision. To better understand the dynamic behavior during the packing process, various dynamic models were proposed to describe the impact process with a combination of quasi-static contact theories and dissipation mechanisms \cite{Marshall14, Li11, Brilliantov96, Brilliantov07, Luding01}. While the dissipation mechanisms account for the dynamic effects such as viscoelasticity, the quasi-static contact theories fundamentally describe the contact between particles, of which Hertz contact theory is the most successful to characterize the relation of applied forces and contact area of non-adhesive particles. However, as particle sizes go down to micron scales, van der Waals adhesion becomes the most important interaction. Several mesoscale models are put forward to describe the effects of van der Waals adhesion on the elastic forces between static contacts of particles, among which the JKR (Johnson, Kendall and Roberts), DMT (Derjaguin, Muller and Toporov) and M-D (Maugis-Dugdale) models are widely accepted ones \cite{Johnson71, Derjaguin94, Liu10}. Recently, Li and Marshall \cite{Li07}, and Marshall \cite{Marshall09} developed a three-dimensional, mesoscopic discrete-element method (DEM) for adhesive micron-sized particles based on the JKR model, which has been successfully applied to dynamic simulations of micron-particle deposition on both flat and cylindrical surfaces with experimental validations \cite{Li11, Yang13}. With this approach of adhesive DEM, both macroscopic and microscopic parameters during the dynamic process can be predicted.

Previous studies using a discrete element method (DEM) have found that the packing fraction decreases for adhesive micron-sized particles in a range of $\phi=0.165-0.622$ with smaller sizes \cite{Yang:2000aa}. A similar result of $\phi \approx 0.2-0.55$ was found for $4-52\mu m$ particles both in simulation and experiment \cite{Parteli:2014aa}. As for other experimental investigations, a random ballistic deposition and fluidized bed technique was used respectively to produce the packing fractions of $\phi=0.15-0.33$ for both uncompressed and compressed samples \cite{Blum06} and of $\phi=0.23-0.41$ with particle diameter of $7.8-19.1\mu m$ \cite{Valverde04}. From these studies, we can conclude that for micron-sized particles, packing density has a positive correlation with particle size. Nevertheless, the effects of other physical parameters, such as velocities, strength of adhesion, frictions etc., on the packing process have little been discussed or analyzed.

In this paper, a prototypical packing system is introduced for the simulation of random loose packings of {\it soft-sphere, non-Brownian}, uniform adhesive particles with a discrete element method. We explore the very low density regime of small particles with van der Waals adhesion interaction by changing physical parameters. The effects of particle velocity, size and work of adhesion on the packing properties are investigated via a dimensionless adhesion parameter. The paper is organized as follows: the adhesive DEM simulation approach based on the JKR theory is described in detail in Section 2.1; the simulation conditions and parameters used in this work are given Section 2.2; and the results and discussions about the effects of the physical parameters on the packing properties are included in Section 3. Section 4 gives the conclusions.

\section{2 Models and Method}

{\it 2.1 Computational Method: Adhesive DEM} \\

In a novel DEM framework specifically developed for adhesive grains \cite{Li11, Marshall14}, both the transitional and rotational motions of each particle in the system are considered on the basis of Newton's second law.
\be
\label{newtonslaw}
\begin{split}
m\frac{d{\bm v}}{dt}={\bm F}_F+{\bm F}_A, \\
I\frac{d{\bm \Omega}}{dt}={\bm M}_F+{\bm M}_A,
\end{split}
\ee
where $\bm{v}$ and $\bm{\Omega}$ are, respectively, velocity and rotation rate of an individual particle, $m$ is the particle mass, and $I=(2/5)mr^2_p$ is the moment of inertia. In a vacuum system that may apply for problems of interstellar dusts, the fluid forces and torques acting on the particle, $\bm{F}_F$ and $\bm{M}_F$, are ignored. $\bm{F}_A$ denotes the collision and the van der Waals adhesion forces in the equation for translational motion. Meanwhile, in the equation for rotational motion, $\bm{M}_A$ denotes the sum of the collision and van der Waals adhesion torques on the particle. They include
\be
\label{forcetorque}
\begin{split}
& \bm{F}_A=F_n\bm{n}+F_s\bm{t}_s,\\
& \bm{M}_A=r_pF_s(\bm{n}\times\bm{t}_s)+M_r(\bm{t}_s\times\bm{n})+M_t\bm{n},
\end{split}
\ee
where $F_n$ is the normal force including adhesively elastic contact force and damping force, $F_s$ is the tangential force due to the sliding friction, $M_r$ is the rolling resistance and $M_t$ is the twisting resistance. $r_p$ is the particle radius. $\bm{n}$, $\bm{t}_s$ and $\bm{t}_r$ are the normal, tangential and rolling direction unit vectors, respectively. \\

{\bf Normal Elastic and Adhesive Forces} \\

The normal force acts in the direction of the unit vector $\bm{n}$ which points parallel to the line connecting the centers of the two particles, denoted by $i$ and $j$, such that $\bm{n}=(\bm{x}_j-\bm{x}_i)/\abs[\bm{x}_j-\bm{x}_i]$. We consider two particles with radii $r_i$ and $r_j$, elastic moduli $E_i$ and $E_j$, and Poisson ratios $\sigma_i$ and $\sigma_j$. An effective particle radius $R$ and effective elastic moduli $E$ and are defined by
\be
\label{elastic}
\frac{1}{R}=\frac{1}{r_i}+\frac{1}{r_j},
\frac{1}{E}=\frac{1-\sigma ^2_i}{E_i}+\frac{1-\sigma ^2_j}{E_j}.
\ee
The particle normal overlap $\delta_N$ is defined by $\delta_N=r_i+r_j-\abs[\bm{x}_i-\bm{x}_j]$ , where $\bm{x}_i$ and $\bm{x}_j$ denote the particle centroid positions. The particle normal adhesive and elastic forces are written together as $\bm{F}_A=F_n\bm{n}$, where a {\it Lennard-Jones} like formula for $\bm{F}_A$ between two particles is proposed (see more details in \cite{Marshall14, Dominik97})
\be
\label{JKR}
\begin{split}
& \frac{F_n}{F_C}=4(\frac{a}{a_0})^3-4(\frac{a}{a_0})^{3/2},\\
& \frac{\delta_N}{\delta_C}=6^{1/3}[2(\frac{a}{a_0})^2-\frac{4}{3}(\frac{a}{a_0})^{1/2}],
\end{split}
\ee
which are in turn based on the JKR theory, as $a^3=\frac{3}{4}R[F_{ne}+3\pi R \omega+\sqrt{6\pi R \omega F+(3\pi R \omega)^2}]/E$, here $a(t)$ is the contact region radius and $\omega$ is work of adhesion due to van der Waals interactions (namely twice the surface energy $\omega=2\gamma$) of two contacting particles \cite{Johnson71}. The JKR model assumes that the adhesive force is only acting inside the contact radius $a$, and results in a larger contact area than the classic Hertzian contact, $a>a_h$. Basically, in contrast to the DMT model (Derjaguin-Mueller-Topprov), the JKR model is appropriate for compliant, adhesive particles for which the particle's Tabor parameter is large than unity, implying the length scale of elastic deformation is larger compared to the length scale of the adhesive force \cite{Marshall14}. Then, the critical force and overlap, $F_C$ and $\delta_C$, and the equilibrium contact radius $a_0$ are given by
\be
\label{critical}
F_C=\frac{3}{2}\pi \omega R,
\delta_C=\frac{a^2_0}{2(6)^{1/3}R},
a_0=(\frac{9\pi \omega R^2}{2E})^{1/3}.
\ee
Typical values of $\omega$ are about $10-30mJ/m^2$, from either measurements \cite{Heim99} or Lifshitz theory's predictions. \\

{\bf Rolling Resistance}\\

The rolling resistance exerts a torque on the particle in the $M_r\bm{t}_r\times\bm{n}$ direction, where $\bm{t}_r$ is the direction of the ``rolling'' velocity. An expression for the rolling displacement of arbitrary-shaped particles is derived by Bagi and Kuhn \cite{Bagi04}. Taking the rate of this expression and specializing to spherical particles of equal size yields an equation for the ``rolling velocity'' $\bm{v}_L$ of particle $i$ as
\be
\label{vl}
\bm{v}_L=-R(\Omega_i-\Omega_j)\times\bm{n}.
\ee
An expression for the rolling resistance torque $M_r$ is postulated in the form
\be
\label{mr}
M_r=-k_r\bm{\xi}\cdot \bm{t}_r,
\ee
where the direction of rolling is $\bm{t}_r=\bm{v}_L/\abs[\bm{v}_L]$ and the rolling displacement is $\bm{\xi}=\int_{t_0} ^t \bm{v}_L(\tau)d\tau$. Rolling involves an upward motion of the particle surfaces in one part of the contact region and a downward motion in the other part. The presence of an adhesive force between the two contacting surfaces thus introduces a torque resisting rolling of the particles. An expression for the rolling resistance in presence of adhesion was derived by Dominik and Tielens \cite{Dominik97, Dominik95}, which yields the coefficient $k_r$ as
\be
\label{kr}
k_r=4F_C(a/a_0)^{3/2}.
\ee
The critical resistance occurs when the rolling displacement magnitude, $\xi=\abs[\bm{\xi}]$, achieves a critical value, corresponding to a critical rolling angle $\theta_{crit}=\xi_{crit}/R$. For $\xi>\xi_{crit}$, the rolling displacement $\bm{\xi}$ in Eq.~\ref{vl} is replaced by $\xi_{crit} \bm{t}_r$. It is noted, according to the measurement by atomic force spectroscopy, $\theta_{crit}$ is around $(0.6-1.0)\%$ \cite{Sumer08}.\\

{\bf Sliding and Twisting Resistance} \\

As aforementioned, both sliding and twisting are relatively rare for small adhesive particles – rolling is generally the preferred deformation mode for agglomerates of adhesive particles \cite{Marshall14, Dominik95}. It is therefore desirable to introduce relatively simple expressions for sliding and twisting resistance in the DEM framework. The standard sliding model for the case without adhesion is the spring-dashpot model proposed by \cite{Cundall79}, for which the sliding force $F_s$ is given by a linear spring-dashpot, $F_s=-k_t\delta_t \cdot\bm{t}_s-\eta_t \bm{v}_s\cdot\bm{t}_s$ ($\bm{t}_s$ is tangential direction), when $\abs[F_s]<F_{crit}$ and by the Amonton friction expression $F_s=-F_{crit}$ when $\abs[F_s]\geq F_{crit}$. The twisting model is given similarly as $M_t=-k_Q\int_{t_0} ^t \Omega_T(\tau)d\tau-\eta_Q \Omega_T$ with $\Omega_T=(\bm{\Omega}_i-\bm{\Omega}_j)\cdot \bm{n}$ the relative twisting rate \cite{Marshall14}. All the model parameters ($k_t$, $\eta_t$, $k_Q$, $\eta_Q$) are chosen based on \cite{Yang13}. Here, a simple model proposed by Thornton \cite{Thornton91a} and Thornton and Yin \cite{ThorntonYin91b}, agreeing reasonably well with experimental data, is introduced. In this model, the only influence of van der Waals adhesion on sliding force is to modify the critical force $F_crit$ at which sliding occurs, which is given by
\be
\label{fcrit}
F_{crit}=\mu_f \abs[F_ne+2F_C],
\ee
where $F_C$ is the critical normal force given in Eq.~\ref{critical} and $\mu_f$ is a friction coefficient that is normally about 0.3. When particles are being pulled apart, the normal force approaches $-F_C$  at the point of separation, at which point the critical sliding force in Eq.~\ref{kr} approaches $\mu_f F_C$. All the values or ranges of $\mu_f$, $\theta_{crit}$ and $F_C$ are selected according to the data from atomic force microscopy measurements \cite{Heim99, Sumer08, Jones04}.

The same model with twisting resistance can be used in the presence of adhesion, with the critical force $F_{crit}$ used to obtain $M_t=\frac{3\pi}{16}aF_{crit}$. For twisting moments with magnitude greater than $M_{t,crit}$, the torsional resistance is given by $M_t=-M_{t,crit}$.\\

{\it 2.2 Simulation Conditions}\\

The generation of the packing starts with the successively random free falling of 1000 uniform spheres with an initial velocity $U_0$ at a height $H$ under gravity. The horizontal deposition plane has two equal edges of length $L$ along with periodic boundary conditions on both directions. A stable packing structure is achieved when all the particles are settled after a time long enough. Here, the fluid effect is filtered out by assuming packing under a vacuum condition. More importantly, the gravitational effect with respect to particle inertia can be neglected when the system satisfies $Fr=U_0/\sqrt{gH}>1$, where $Fr$ is the {\it Froude} number (ratio of inertia to gravity). To precisely characterize the gravitational effect, we define the relative velocity increment as $\Delta U=(U-U_0)/U_0$, where $U=\sqrt{U_0^2+2gH}$ is the final velocity. For all runs in the numerical simulations, we ensure that $\Delta U$ is less than $4\%$. Thus the gravitational acceleration during the dynamic deposition process can be ignored, indicating that the deposition velocity can be treated as the same with the initial inlet velocity. After the packing is formed, however, gravity still works and can't be neglected. As a primary concern of our work, interstellar dust particles always transport with a relatively large velocity before they form large aggregates. Therefore, the adhesive packings simply arise due to the competition between the particle inertia and particle-particle interactions (e.g., adhesion, elasticity and frictions). Most importantly, the negligible gravitational effect distinguishes our system from that of \cite{Yang:2000aa, Parteli:2014aa}, which generate particles randomly in a box without touching each other and wait them to deposit due to gravity.

Before the simulation, a sensitivity analysis between the cases $L=20r_p$ and $L=40r_p$ was conducted with parameters in Table~\ref{table1}. As also shown in Table~\ref{table1}, the difference in $\phi$ between the cases $L=20r_p$ and $L=40r_p$ is negligible, indicating the $L=20r_p$ is large enough to reproduce bulk properties. Then, we set $L=20r_p$. The physical and geometrical parameters used in the DEM simulations are listed in Table~\ref{table2}. In addition to changing particle size, we applied different work of adhesion and initial velocities to explore their effects on adhesive packings.

\begin{table}[h]
\small
  \caption{\ Sensitivity analysis between cases of $L=20r_p$ and $L=40r_p$.}
  \label{table1}
  \begin{tabular*}{0.5\textwidth}{@{\extracolsep{\fill}}ccccccc}
    \hline
    $r_p$ & $L$ & $U_0$ & $\omega$ & $N$ & Volume & Coordination \\
    $(\mu m)$ & $(\mu m)$ & $(m/s)$ & $(mJ/m^2)$ &  & Fraction & Number \\
    \hline    
    1 & 20 & 0.5 & 30 & 1000 & 0.155 & 2.24 \\
    1 & 40 & 0.5 & 30 & 4000 & 0.157 & 2.25 \\
    5 & 100 & 0.5 & 30 & 1000 & 0.265 & 2.80 \\
    5 & 200 & 0.5 & 30 & 4000 & 0.270 & 2.76 \\
    10 & 200 & 0.5 & 30 & 1000 & 0.372 & 3.18 \\
    10 & 400 & 0.5 & 30 & 4000 & 0.375 & 3.19 \\
    \hline
  \end{tabular*}
\end{table}

\begin{table}[h]
\small
  \caption{\ Parameters used in DEM simulations.}
  \label{table2}
  \begin{tabular*}{0.5\textwidth}{@{\extracolsep{\fill}}ccc}
    \hline
    Physical Parameters & Values & Units \\
    \hline    
    Particle Number($N$) & 1000 &  \\
    Particle Radius($r_p$) & 1,5,10,50 & $\mu m$ \\
    Particle Density($\rho_p$) & 2500 & $kg/m^3$ \\
    Work of Adhesion($\omega$) & 30,20,10,5,1,0.1 & $mJ/m^2$ \\
    Characteristic Length($L$) & $20r_p$ & $\mu m$ \\
    Particle Injection Height($H$) & $4\times L$ & $\mu m$ \\
    Gravity Acceleration($g$) & 9.81 & $m/s^2$ \\
    Deposition Velocity($U_0$) & 0.5-10 & $m/s$ \\
    \hline
  \end{tabular*}
\end{table}

\section{3 Results and Discussions}

{\it 3.1 Volume Fraction} \\

\begin{figure}
\begin{center}
\includegraphics[width=7.5cm]{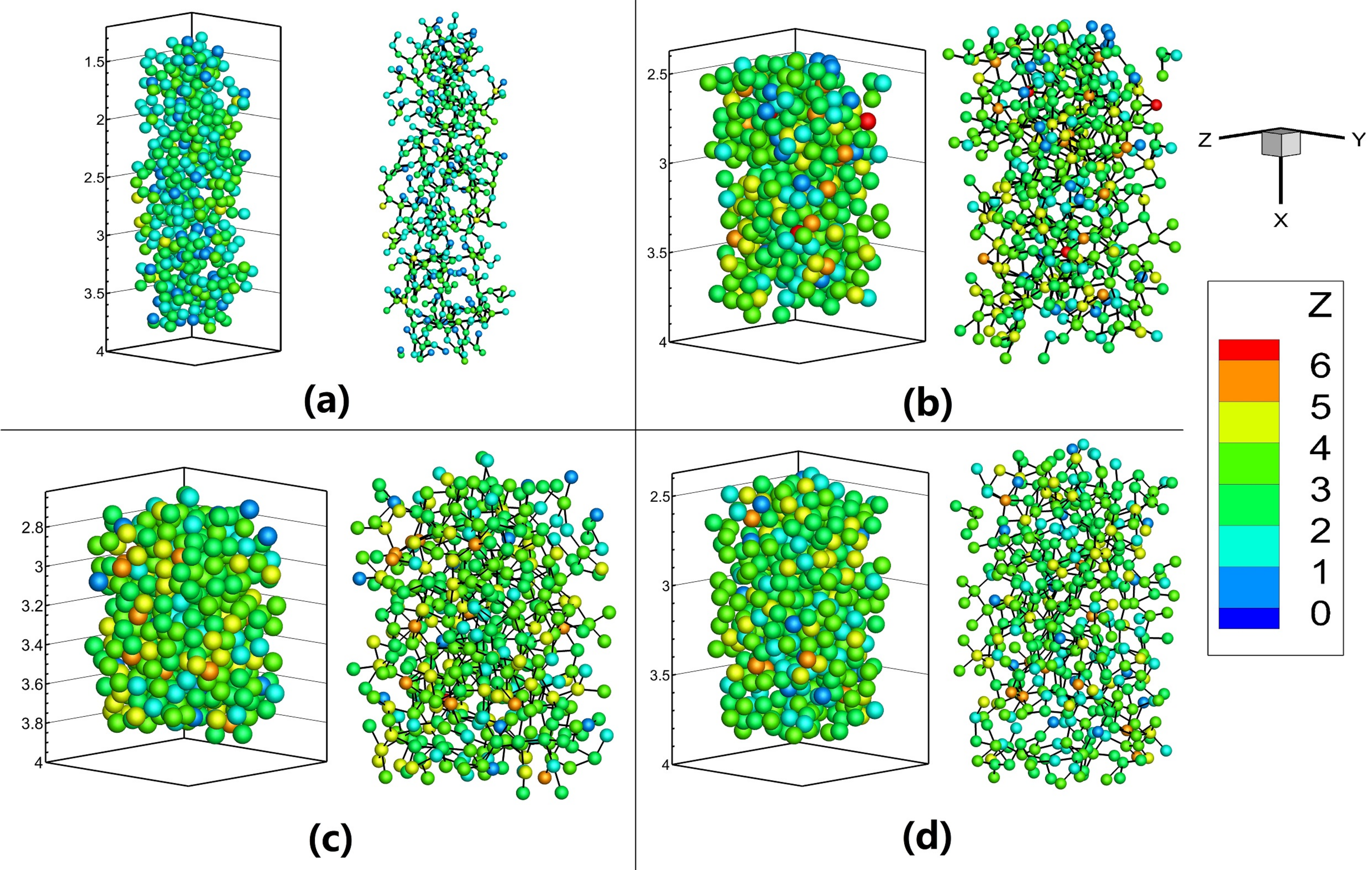}
\caption{\label{Fig_structure}(Colors online) Typical packing structures with varying physical parameters. (a) stands for a basic loose packing with $r_p=1\mu m$, $U_0=1m/s$ and $\omega=30mJ/m^2$. The other three plots change only one physical parameter for each, which are $U_0=2m/s$ in (b), $r_p=5\mu m$ in (c) and $\omega=10mJ/m^2$ in (d) with other parameters fixed respectively. The right part in each plot shows the connecting network of packing structures. Different color represents different coordination number $Z$. }
\end{center}
\end{figure}

Typical packing structures and connecting networks of loose and dense packings are shown in Fig.~\ref{Fig_structure}. The volume fraction is calculated from the vertically middle part of the packing ($0.3h\leq X \leq 0.8h$, with $h$ as packing height), avoiding the so-called wall effect from both bottom and top of the packing structure. It has been well accepted that volume fraction increases with the increment of particle size \cite{Yang:2000aa, Parteli:2014aa, Valverde04}. However, from Fig.~\ref{Fig_structure} we can see that not only the particle size ($r_p$) but also the deposition velocity ($U_0$) and work of adhesion ($\omega$) both significantly affect mesoscopic packing structures. The increased $U_0$ or decreased $\omega$ will result in a relatively dense packing, which is primarily due to the competition between particle inertia and adhesion that will be discussed below. Also from Fig.~\ref{Fig_structure}, in a very loose packing the connecting network turns out to be like chains with only two contacts for most of the particle while for a dense one there are more small loops in the network.

\begin{figure}
\begin{center}
\includegraphics[width=7.5cm]{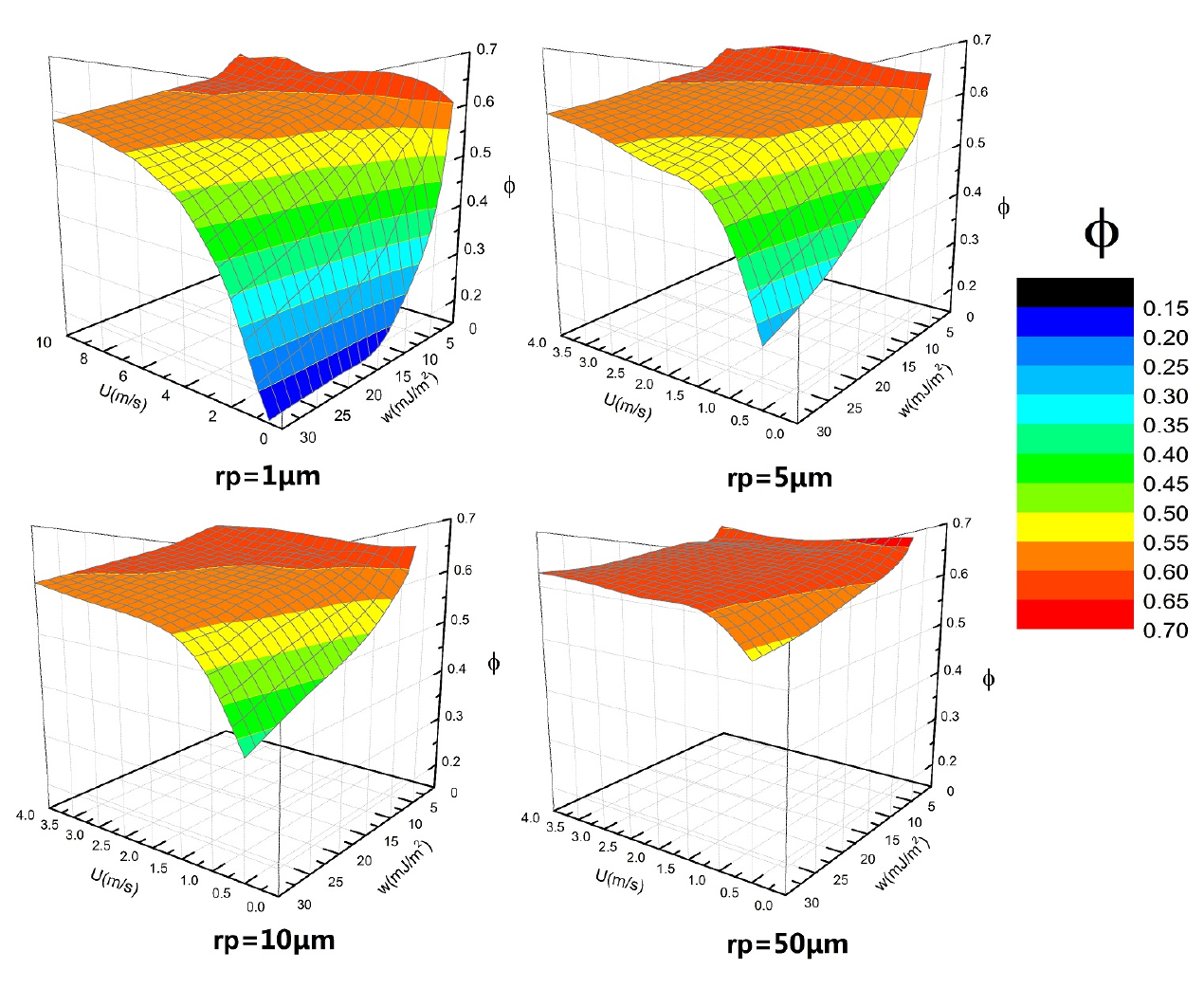}
\caption{\label{Fig_phiruw}(Colors online) 3D plot of volume fraction with different velocities and work of adhesion under the conditions of different sizes of particles. Left top ($r_p=1\mu m$), right top ($r_p=5\mu m$), left bottom ($r_p=10\mu m$), right bottom ($r_p=50\mu m$). }
\end{center}
\end{figure}

We then figure out the volume fraction of all the simulation conditions under different sizes of particles ($r_p$), deposition velocities ($U_0$) and work of adhesion ($\omega$). It is seen from the four sub-plots in Fig.~\ref{Fig_phiruw} that, with increased deposition velocity, the volume fraction increases and then stays at a value approaching 0.64, which is the widely accepted as the upper limit of random close packing ($\phi_{RCP}$) with uniform sphere particles. However, the extent of the increment is somewhat different under variation of $r_p$ or under variation of $\omega$. For instance, for a relatively small particles ($r_p=1\mu m$), it is interesting that the very loose packing ($\phi=0.155$) can be achieved at a low velocity ($U_0=0.5m/s$) with strong adhesion ($\omega=30mJ/m^2$). When the deposition velocity grows to $10m/s$, the volume fraction reaches 0.582 with an increment of $\sim 0.427$. However, when $\omega$ goes down to $0.1mJ/m^2$ (nearly non-adhesive), the volume fraction hardly changes with $U_0$. On the other extreme, for much bigger particles ($r_p=50\mu m$), despite the changes of either $U_0$ or $\omega$, the volume fraction seems to converge to a horizontal plane, which means the effects of both adhesion and particle inertia on the volume fraction is sufficiently small for big grains.

\begin{figure}
\begin{center}
\includegraphics[width=7.5cm]{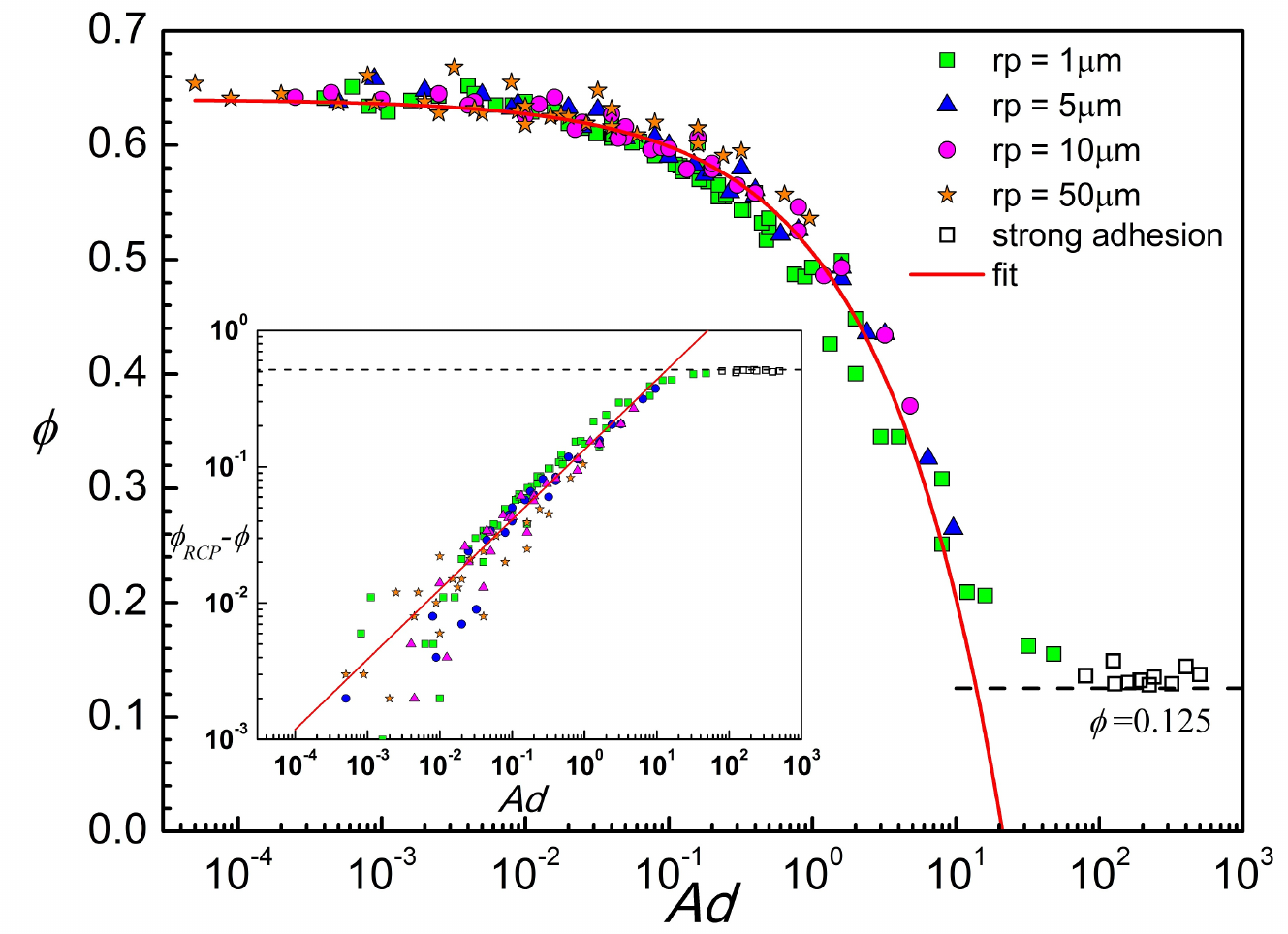}
\caption{\label{Fig_phiAd}(Colors online) Packing volume fraction as a function of adhesion parameter. The black open squares are the cases with strong adhesion. The red line is a fitting of $\phi$ versus $Ad$. }
\end{center}
\end{figure}

A dimensionless adhesion parameter $Ad=\omega/2\rho_p U_0^2R$, which interprets the balance between the interparticle adhesion and the particle inertia, has been successfully applied to physically lump together the effects of particle size, velocity and work of adhesion \cite{Li07, Liu15}. Fig.~\ref{Fig_phiAd} further extends the packing properties to both higher and lower $Ad$ conditions than those in \cite{Liu15}. In the case of $Ad<1$, the volume fraction varies smoothly from $\sim 0.50$ (close to RLP) to $\sim 0.64$ (RCP), with particle inertia dominating over adhesion. However, when $1<Ad<20$, an adhesion-controlled regime is observed, where the volume fraction decreases exponentially with increasing $Ad$. Furthermore, with very high $Ad(>20)$, there seems to be a lower limit of the volume fraction which determines a physical packing. The work of adhesion is increased to $\omega=50-250mJ/m^2$ while the velocity is decreased to $0.2m/s$ with fixed particle size $r_p=1\mu m$ in the simulations in order to achieve packings with very high $Ad(=80-500)$. As indicated by the black open squares in Fig.~\ref{Fig_phiAd}, we can see that the volume fraction no longer decreases with $Ad>100$ and the lowest volume fraction is 0.128, which is very close to the well-known lower bound of saturated sphere packings in $d$ dimensions $\phi=1/2^d$ \cite{Torquato06, Liu15}.

Regarding the densification with the decrease of $Ad$, two potential ways of compaction may be the causes: gravitational (or other external forces) compaction and particle inertial compaction. Following the work of \cite{Blum06}, the pressure caused by these two compactions can be defined as $p_g=\frac{4}{3}\rho_p gr_p$ and $p_i \approx \frac{1}{4}\rho_p U_0^2$. With the parameters given in Table~\ref{table2}, we get $p_g=(0.133\sim 1.63)Pa$ and $p_i \approx (156.25\sim 6.25\times 10^4)Pa$. Thus, the compaction caused by gravity is negligible and particle inertial compression is the most important way of densification, which further approves our primary assumption that gravitational effect can be negligible. Furthermore, from the definition of $Ad$ we can see that the compression resulted from particle inertia has already been considered and characterized with respect to adhesion. When particles are being packed, the van der Waals adhesion forces tend to attract particles and make them stick together. On the other hand, the particle inertia which has a quadratic correlation with particle velocity will urge them to move and impact with other particles. If adhesion is stronger than particle inertia, particles are more likely to be caught at the incipient impacts and hardly move or roll. Thus a loose packing structure is easier to form. With the increase of particle size or velocity, particle inertia will become much stronger than adhesion. As a consequence, more collisions will take place and particles tend to explore more phase space and form a denser packing structure. Note that in some previous studies the volume fraction is also related to the force ratio, or the so-called {\it Bond} number, which is defined as $Bo=f_c/f_e$ \cite{Dong06, Singh14}. $f_c$ is the attractive cohesive forces which may come from van der Waals interactions, liquid bridge cohesion and so on while $f_e$ represents the external driven forces that could be gravity \cite{Dong06} or average force due to external compression \cite{Singh14}. However, as discussed above, the balance of particle inertia and adhesion dominates our packings instead of gravity, meaning that we can still achieve different packing structures by changing velocities while $Bo$ is unchanged. Therefore, for our inertia-adhesion controlled system, $Ad$ is more appropriate to characterize the mechanism than $Bo$, which is often applied in gravity-cohesion controlled system.

As mentioned above, the volume fraction seems to decrease exponentially with the increasing $Ad$ under conditions of $Ad<20$, such that an approximately liner fit can be well obtained by using the log-plot of  $\phi_0-\phi \sim Ad$, except those points with a certain fluctuation under extra lower $Ad$ parameter (e.g $Ad<0.01$). $\phi_0$ is an asymptotic packing fraction value without adhesion effect. It may be close to the RCP limit of $\phi_{RCP}=0.64$ since the inertial compression exceeds the frictional force in current particulate system. Therefore, a power-low relationship is proposed as
\be
\label{phi}
\phi_{RCP}-\phi=\alpha Ad^\lambda ,
\ee
The parameters $\alpha$ and $\lambda$ can be obtained as $\alpha=0.134\pm0.005$ and $\lambda=0.513\pm0.016$ by a best fitting. Therefore the relationship is $\phi_{RCP}-\phi=0.134 Ad^{0.513}$, which is also plotted in Fig.~\ref{Fig_phiAd} as a red line. The inset of Fig.~\ref{Fig_phiAd} shows the log-log plot of this relation, from where we can see the linear part of $(\phi_0-\phi) \sim Ad$ along with the transition to the lower limit when $Ad>20$. Basically, this relationship connects the lowest volume fraction and the well-known random close packing limit smoothly via a simple parameter $Ad$, though the intrinsically underlying physics behind this relationship still need to be further explored.

Here more discussion is essential with respect to the scaling of $Ad$ as it indicates a universal prediction that the limit of small $Ad$ will lead to only RCP instead of RLP. From the definition we know that $Ad$ is inversely proportional to the square of particle velocity which will result in the compression caused by particle inertia. Thus, the very small values of $Ad$ are mostly obtained with large velocities instead of large particle sizes or weak work of adhesion and the compression will make the packings denser and approach RCP. However, RLP is indeed reached in our results, which is only in a small range of $Ad$ (around 0.1). This contrasts with the previous friction-dependent RLP results since friction is finite and fixed in this work. With the tuning of $Ad$, transition from RLP to RCP can also be realized without changing friction. As the effect of friction on adhesive packings is not a main concern of this study, we leave it to the future work. On the other hand, all the results with $Ad>1$ are below RLP, which approve the critical value of $Ad=1$ that distinguishes adhesive packings from adhesive-less ones. Therefore we can define four regimes from the results in Fig.~\ref{Fig_phiAd}: RCP regime with $Ad<\sim 0.01$, RLP regime with $\sim 0.01<Ad<1$, adhesion regime with $1<Ad<20$ and an asymptotic regime with $Ad>20$, where the packings begin to approach the lower limit $\phi=1/2^3$ that is not determined by any physical parameters like adhesion, inertia or friction. \\

{\it 3.2 Coordination number} \\

\begin{figure}
\begin{center}
\includegraphics[width=7.5cm]{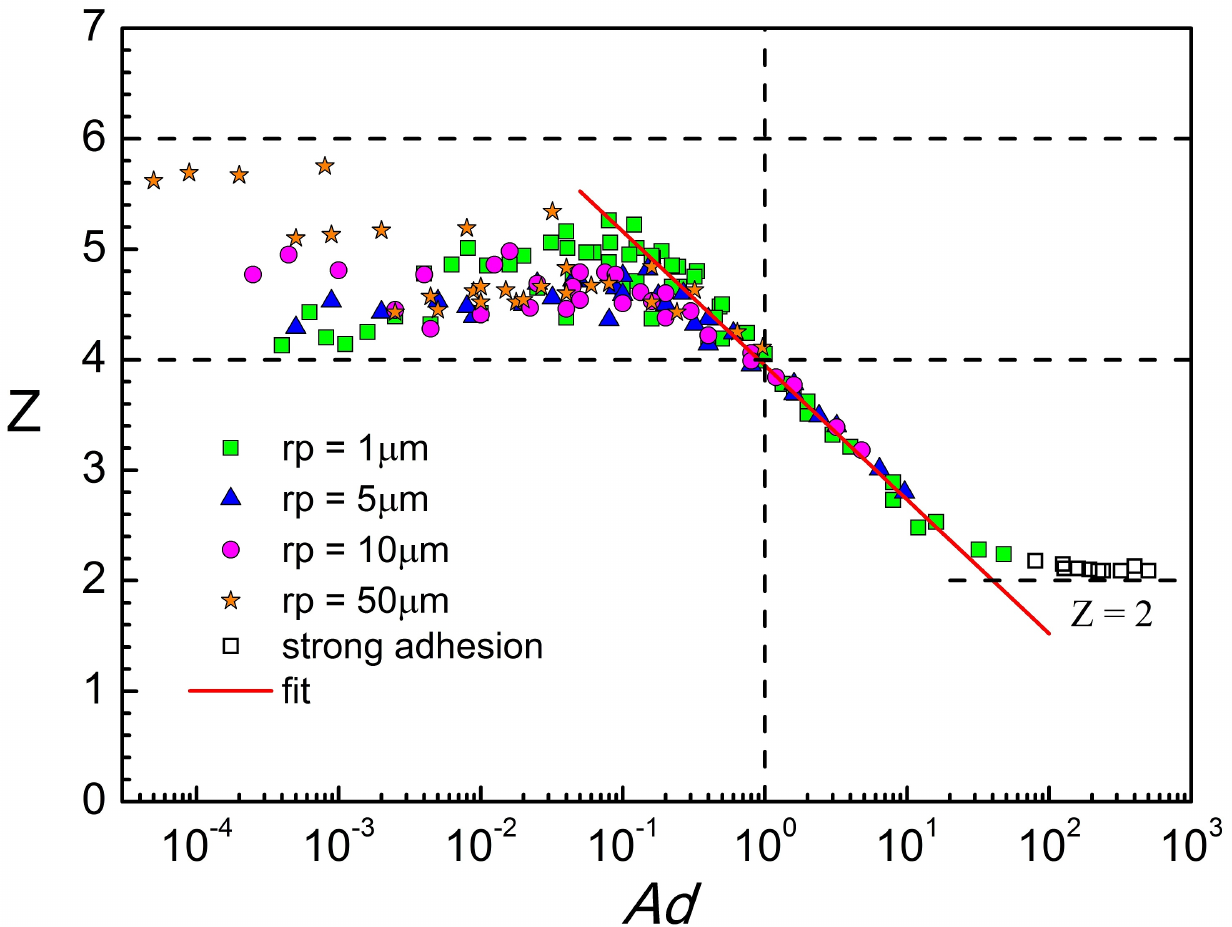}
\caption{\label{Fig_ZAd}(Colors online) Coordination number as a function of adhesion parameter. The black dash lines represent $Z=4$, $Z=6$ and $Ad=1$ respectively. The red line is a fitting of $Z$ versus $Ad$. }
\end{center}
\end{figure}

Coordination number ($Z$), defined as the number of neighbours touching a given particle, is introduced to characterize the packing structures. We calculate $Z$ by judging whether the distance of the centers of two particles is smaller than the sum of their radius. Thus this $Z$ is the geometrical coordination number since it is defined with a criterion of geometric relation that doesn't include any forces. The mean coordination numbers of all the simulation cases are plotted in Fig.~\ref{Fig_ZAd} as a function of $Ad$. We can see that for $Ad<1$ the coordination number lies indeed within the isostatic limits $d+1\leq Z \leq 2d$ where $d$ is the spatial dimension. As mentioned in the computational method section, all of our packings remain frictional with fixed friction coefficient $\mu_f=0.3$ so that the coordination number $Z$ does not reach the isostatic limit $Z=6$ (of frictionless particles) at $Ad<1$. For micron-sized particles, friction is often coupled with adhesion and is complicated to single out. Thus we don't pay much attention to the effect of friction on adhesive packings for the present work. However, for $Ad>1$ the coordination number falls on a unique curve, analogous to the $\phi$ dependence. We figure out a fitting line of $Z$ versus $Ad$ as $Z=3.915-1.113log(Ad)$. Similarly, there also seems to be an asymptotic value of coordination number when $Ad>20$, where the lowest $Z$ reached in our simulations is $Z=2.09$ with $Ad>100$. For these very loose packings, as shown in Fig.~\ref{Fig_structure}, the packing structure is mainly composed of particle chains with coordination number equal to 2. However, to expand the whole packing structure some cross-links in the network are essential, which will result in a local coordination number at least of 3 or 4. Meanwhile, there are also some particles hanging at the end of the chains leading to a local coordination number of 1. Therefore, the average coordination number of the packing is going to be around 2. Regarding the stability of these particles with two or less neighbors, the existence of adhesion still guarantees the mechanical equilibrium. The negligible gravity can be balanced through adhesion and friction and thus these particles can be easily stabilized.

\begin{figure}
\begin{center}
\includegraphics[width=7.5cm]{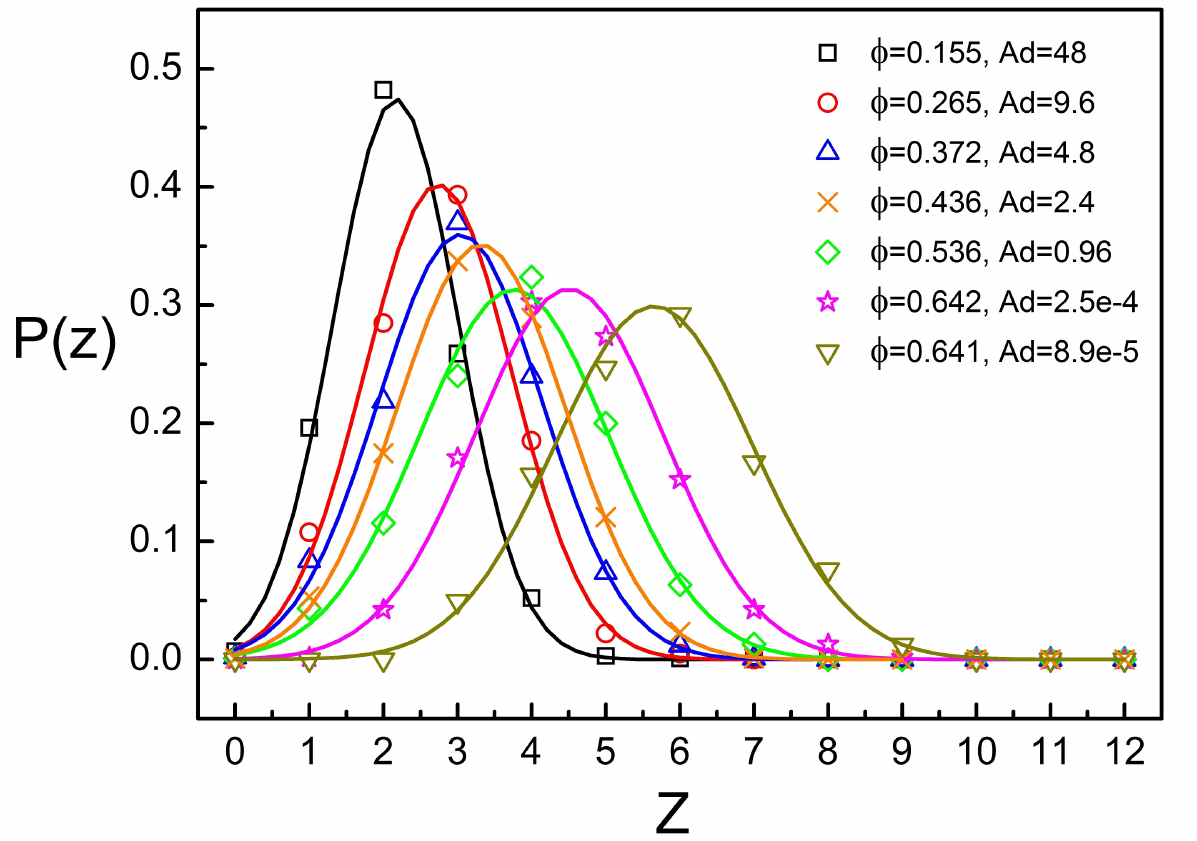}
\caption{\label{Fig_disZ}(Colors online) Distribution of coordination number for different volume fraction and $Ad$. The solid lines are normal distribution fit. }
\end{center}
\end{figure}

Apart from the $Z\sim Ad$ curve, we also investigate the distribution of coordination number. Typical cases are picked among all the simulations to display the common features. As shown in Fig.~\ref{Fig_disZ}, in the packing with the lowest density, most of the particles have coordination number in the range of $1\sim 4$ with a mean value 2.24. With the increase of volume fraction, the distribution of $Z$ shifts to the right (larger values) and becomes wider. It's also noted that the distribution of coordination number is symmetric and shapes like normal distribution. Hence we made a normal distribution fit and the results are listed in Table~\ref{table3}. As we can see, the normal distribution fits well with the data, indicating a ubiquity of normal distribution of coordination number in random packings of uniform spherical particles. \\

\begin{table}[h]
\small
  \caption{\ Average packing properties of different $Ad$ along with a normal distribution fit of $Z$.}
  \label{table3}
  \begin{tabular*}{0.5\textwidth}{@{\extracolsep{\fill}}cccccccc}
    \hline
    Case & a & b & c & d & e & f & g \\
    \hline    
    $Ad$ & 48 & 9.6 & 4.8 & 2.4 & 0.96 & 2.5e-4 & 8.9e-5 \\
    $\phi$ & 0.155 & 0.265 & 0.372 & 0.436 & 0.536 & 0.642 & 0.641 \\
    $Z$ & 2.24 & 2.80 & 3.18 & 3.49 & 4.11 & 4.77 & 5.69 \\
    Fit of $Z$ & 2.166 & 2.738 & 3.035 & 3.319 & 3.761 & 4.505 & 5.648 \\
    Variance & 0.841 & 0.992 & 1.109 & 1.136 & 1.274 & 1.272 & 1.334 \\
    \hline
  \end{tabular*}
\end{table}

{\it 3.3 Radial distribution function}\\

\begin{figure}
\begin{center}
\includegraphics[width=7.5cm]{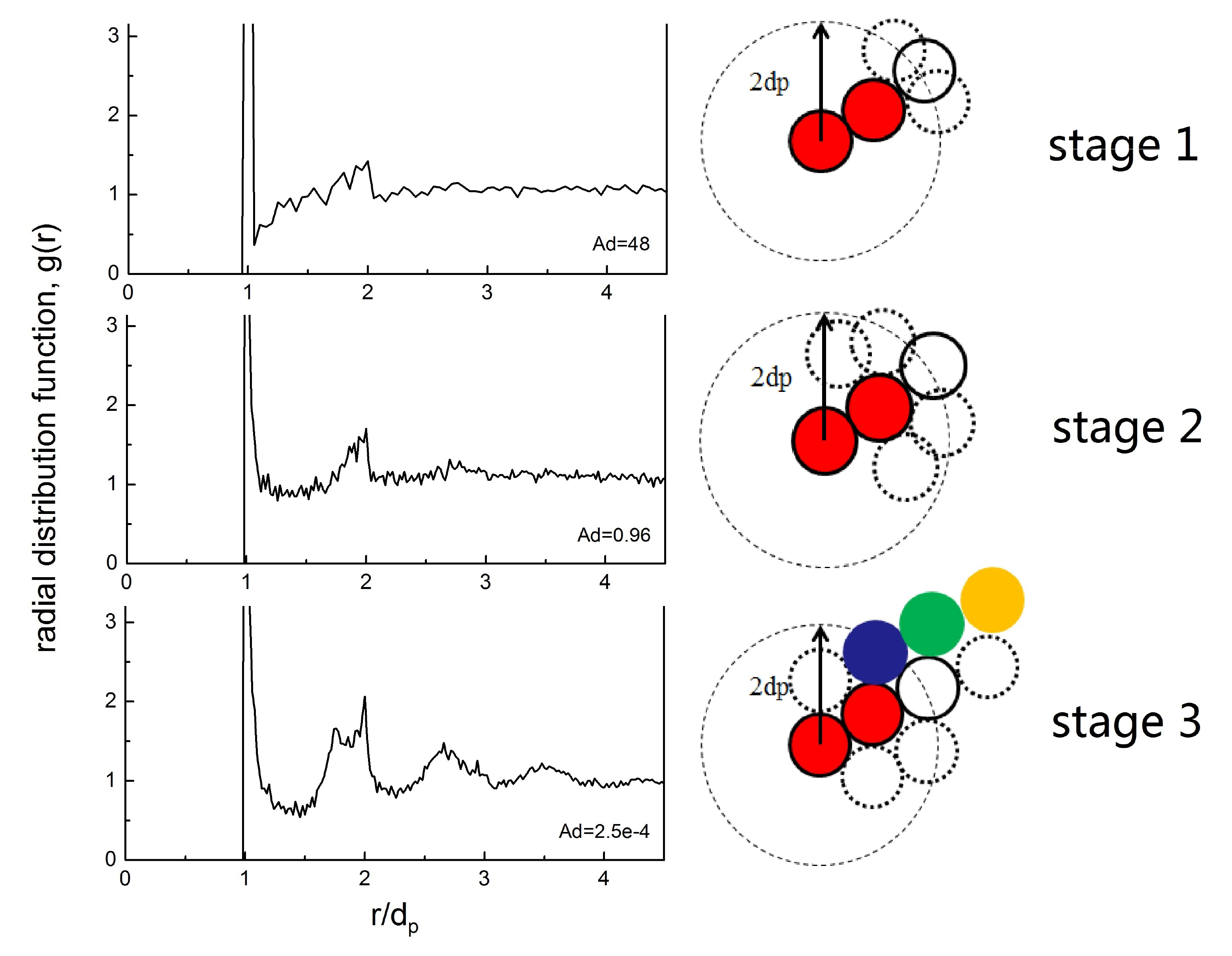}
\caption{\label{Fig_gr}(Colors online) Radial distribution function for packings with different $Ad$. }
\end{center}
\end{figure}

Radial distribution function $g(r)$, which describes the probability of finding a particle at a distance $r$ from a given one, has been widely used to depict the interparticle correlations and characterize the packing density as well as the coordination number. Fig.~\ref{Fig_gr} shows the radial distribution function $g(r)$ of typical packings listed in Table~\ref{table3} with various $Ad$. For loose packings with $\phi<0.4$, we apply $dr=0.05d_p$ to calculate the $g(r)$ while $dr=0.02d_p$ for dense packings with $\phi>0.4$. There seems to be three stages of the evolution of the $g(r)$ with respect to the increasing of volume fraction. {\it (i)} Firstly, the $g(r)$ of very loose packings (about $\phi<0.3$) drops below $g(r)=1$ instantly after the first peak and then gradually rises to the second peak which is inconspicuous here but implies the second contact shell of typical structural features of random packings of granular spheres. The feature that the lowest valley of $g(r)$ is close to $r=d_p$ indicates a very low probability to find a particle just outside the first contact shell. The gradual increase from the valley to the second peak directly describes the greatest possibility to find a particle at the distance $r=2d_p$, accounting for the chainlike structure of very loose packings. After the second peak the $g(r)$ becomes almost flat, indicating a uniform probability to find a particle outside the distance of $r>2d_p$. {\it (ii)} When the volume fraction grows larger (about $0.3<\phi<\phi_{RLP}$), no valleys with $g(r)<1$ appear and the second peak becomes more and more evident along with a flat plateau after $r>2d_p$ as well. This characteristic reflects the uniform probability of finding particles between $r=d_p$ and $r=2d_p$ for typical random loose packings besides the strong peak at $r=2d_p$ which grows up with the increasing of packing density. {\it (iii)} Further, as packing becomes denser, approaching the random close packing, valleys of $g(r)<1$ appear again together with other peaks at typical distance $r=\sqrt{3}d_p$, $r=\sqrt{7}d_p$ and $r=\sqrt{13}d_p$. The valley with the peak at $r=\sqrt{3}d_p$ depicts the adjacent local ordering that four particles form two edge-sharing in-plane equilateral triangles, which is part of the densest packing structure. Despite the ordering at farther distance represented by other peaks, the not-sharp and wide peaks still can't make the packings reach the densest structure like face center cubic (FCC). In fact, it's always improbable to obtain FCC with random packings since the possibility of achieving both the local and global ordering structures of FCC at the same time is almost zero. Only by sufficient compression or vibration can one get a rather dense packing which might be beyond the random close packing. However, no matter how dense the packing is, the $g(r)$ should resemble that of stage 3 since $g(r)$ describes the spatial correlations which are defined geometrically. Despite that the features of $g(r)$ found in our simulations are in consistent with results from \cite{Yang:2000aa}, nonetheless, it should be noted that $g(r)$ only has relation with volume fraction which can be tuned via not only particle size \cite{Yang:2000aa} but also $Ad$. A broad range of volume fraction and various packing structures can be achieved by tuning $Ad$. The three stages of $g(r)$ describe the evolution of packing density and structure of adhesive micron particles, which bridges the loosest and densest random packings through a simple parameter $Ad$. \\

{\it 3.4 Force distribution and mechanical equilibrium}\\

\begin{figure}
\begin{center}
\includegraphics[width=7.5cm]{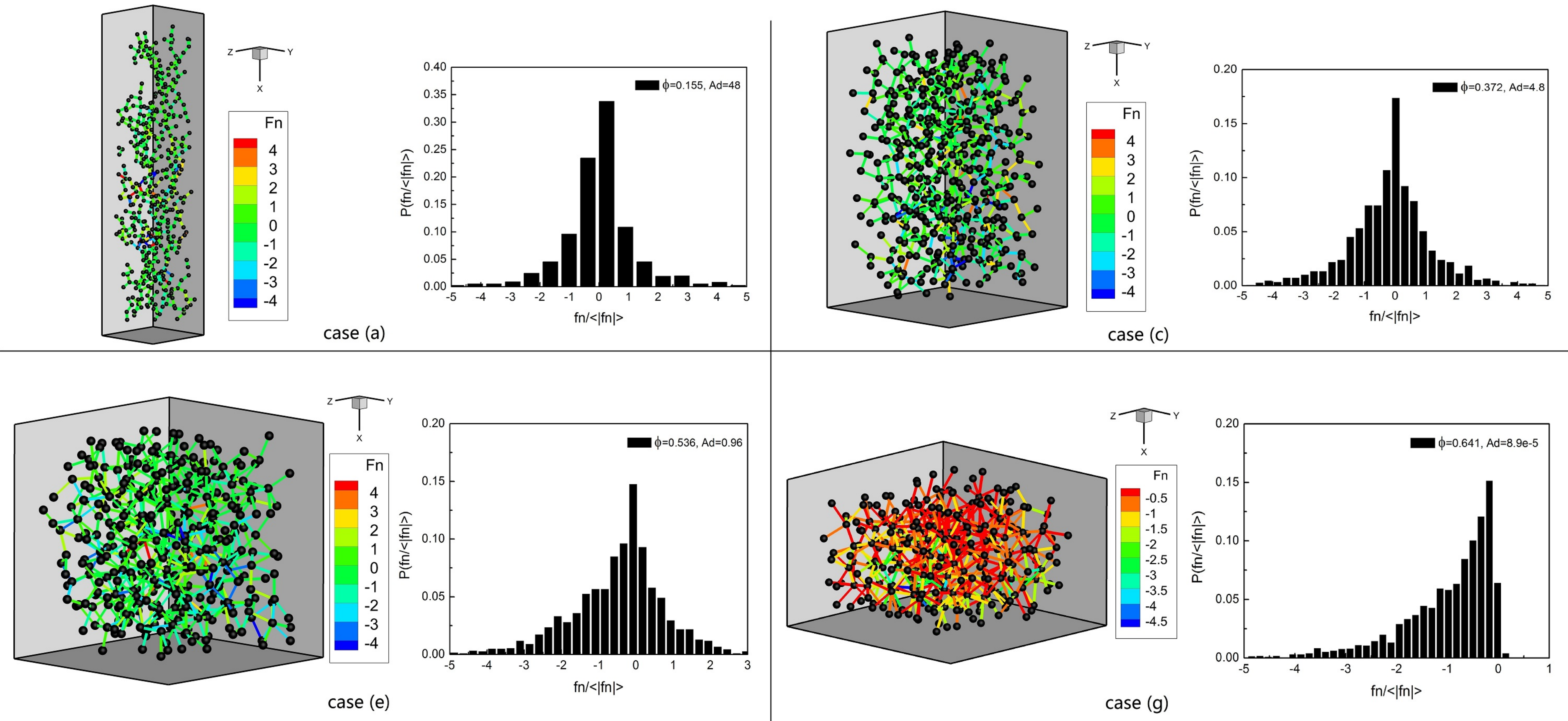}
\caption{\label{Fig_signedf}(Colors online) Signed normal force network and distribution of typical packings of adhesive particles. Negative values are repulsive while positive are attractive. All the forces are normalized with the mean value of the magnitude of normal forces. Case (a)(c)(e)(g) are consistent with the cases listed in Table~\ref{table3}. }
\end{center}
\end{figure}

For loose packings of adhesive particles, especially for the very loose packings with mean coordination number $2\sim 4$, it is of great interest how the particles get mechanical equilibrium, or what the force distribution is like. To investigate these properties, we measure the force of every contact. Fig.~\ref{Fig_signedf} displays the signed normal force network and distribution of four typical packings we have achieved, which are also among the cases listed in Table~\ref{table3}. Here the signed forces means they could be both attractive and repulsive forces between two contact particles. Forces signed negative are repulsive while the positive forces are attractive. It can be directly inferred from Fig.~\ref{Fig_signedf} that both attractive and repulsive forces appear for all the packing densities. This property distinguishes the packing of adhesive particles from that of adhesive-less granular matters where all the forces are repulsive. Furthermore, for low packing densities ($\phi=0.155,0.372,0.536$) the number of attractive forces seems to be the same as repulsive forces, implying a symmetric distribution of signed normal force. With the relatively high packing density ($\phi=0.641$), the number of attractive forces gradually decreases and the force network resembles the granular matter behaviour (see the right bottom of Fig.~\ref{Fig_signedf}). The cause of this behavior can be related to the dependence of $\phi$ on $Ad$. From Fig.~\ref{Fig_phiAd} we know $\phi$ increases with the decreasing of $Ad$, which represents the strength of adhesion. A relative decreasing of $Ad$ will then lead to the reduction of attractive forces.

\begin{figure}
\begin{center}
\includegraphics[width=7.5cm]{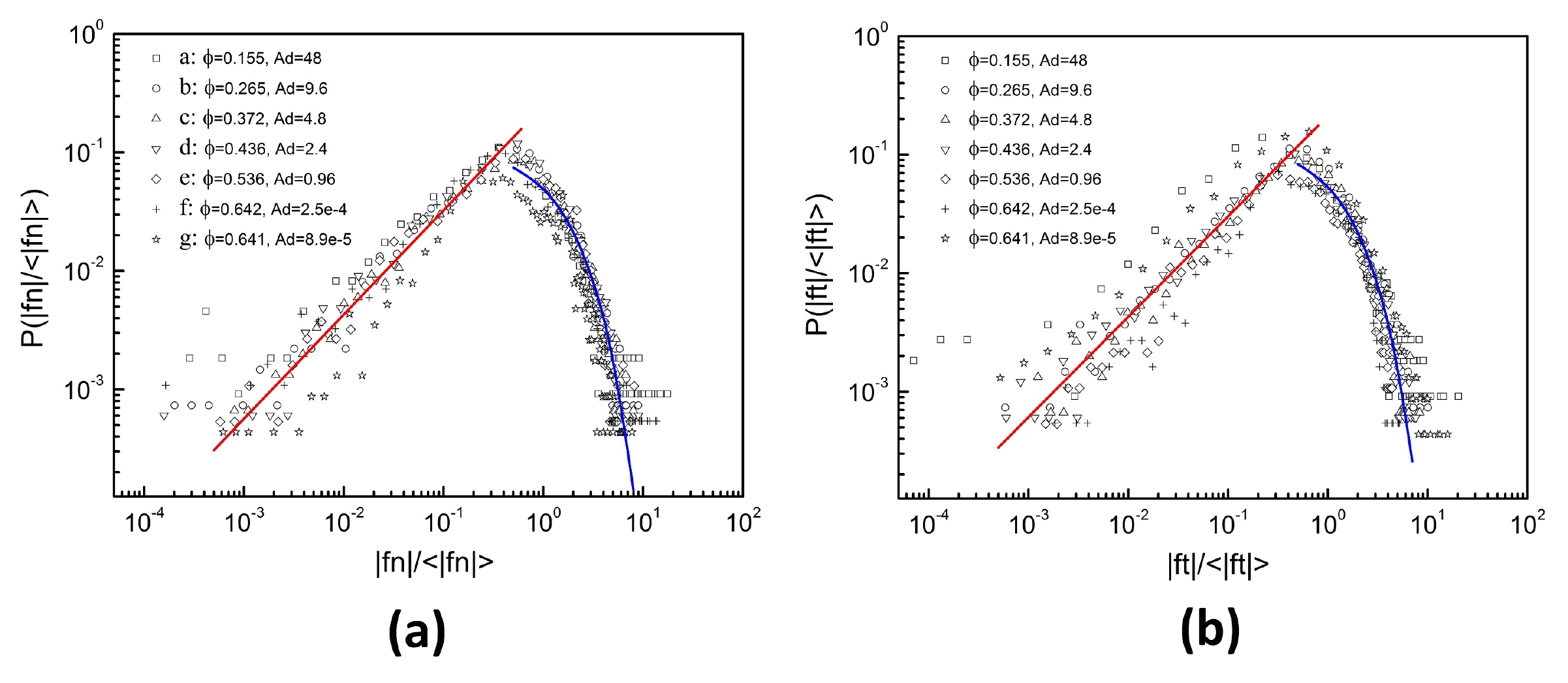}
\caption{\label{Fig_unsignedf}(Colors online) Unsigned normal (a) and tangential (b) force distribution of typical packings of adhesive particles. Case (a)~(g) in the legend are consistent with the cases listed in Table~\ref{table3}. }
\end{center}
\end{figure}

Another important property of the force network is that, no matter what the packing density is, most of the forces are around zero which identifies the small forces behavior. In order to study this behavior, we measure the unsigned force, namely the magnitude of the attractive and repulsive forces. Fig.~\ref{Fig_unsignedf} shows the distribution of both unsigned normal and tangential forces of the typical adhesive packings that listed in Table~\ref{table3}. The plateau at the biggest forces and the deviation of the smallest forces are believed to be caused by the finite size effect, where the probability of finding the biggest or smallest force should be around $1/NZ$ ( $10^-3$ to $3\times 10^-4$ here). We find the similar scaling $P(f)\sim f^\theta$ for small forces as well as the power law $P(f)\sim \exp^{-\beta f}$ for big forces \cite{Mueth98, Lerner13, Bo14}. However, the exponents are $\theta=0.879$, $\beta=0.839$ for normal forces while $\theta=0.848$, $\beta=0.888$ for tangential forces, respectively. These exponents are distinct from the reported values of $\theta \approx 0.2 \sim 0.5$ and $\beta \approx 1.0 \sim 1.9$ for adhesive-less packings \cite{Mueth98, Lerner13, Bo14}. The exponential decay of big forces for non-adhesive packings can be predicted by the ``q model'', in which the total weight on a given particle is transmitted randomly to its nearest underneath neighbours with different fractions $q_{ij}$ of load on each particle \cite{Liu95}. With the sum of the fraction equaling to one, the sum of all the force components in the vertical direction is conserved. Nevertheless, this q model is still too simple to account for the force behaviour of adhesive packings for two reasons. One is that only the vertical forces are considered. The other is that the existence of adhesion weakens the load that the underneath particles carry from the upper ones. Thus it might lead the fraction $q_{ij}$ to be negative, where the forces between two contact particles are attractive. But we still speculate that the adhesion will not make much effect on the distribution of big forces. On the other hand, for small forces, it is reported that they have a close relation with the stability of packing. Contacts carrying small forces which are repulsive for granular matters are easy to break. However, for adhesive particles, the process of necking will take place and a critical pull-off force $F_C$ needs to be reached to separate two contact particles. Thus contacts with small forces in adhesive packings are relatively more stable, no matter attractive or repulsive. Regarding the exponents, we conjecture that the very small and very big repulsive forces in granular packing will not likely exist in the presence of adhesion. The very small repulsive forces could be enhanced while the very big ones will be weakened by attractive adhesion force, which will result in the shrink of the force distribution. Therefore the exponent $\theta$ of small forces will increase a little and $\beta$ of big forces will decrease. However, this is just a qualitative analysis where there have been few data or theory to support. More insight from both experiment and theoretical efforts is left to future work.

For granular matters without adhesion, at least three particles (in 3D) are usually needed to support one and make it mechanical equilibrium. However, for the very loose packings of adhesive particles, many particles only have one or two neighbours which are not able to achieve mechanical equilibrium in terms of granular matters. Indeed, these particles can still be mechanically stabilized for two reasons. The first is the adhesive attractive force caused by van der Waals interactions that are especially prominent for micron-sized particles. The other one is friction as well as the negligible gravity compared with van der Waals force. Next we will conduct the mechanical equilibrium analysis and explain qualitatively why adhesive particles can be stabilized more easily.

\begin{figure}
\begin{center}
\includegraphics[width=7.5cm]{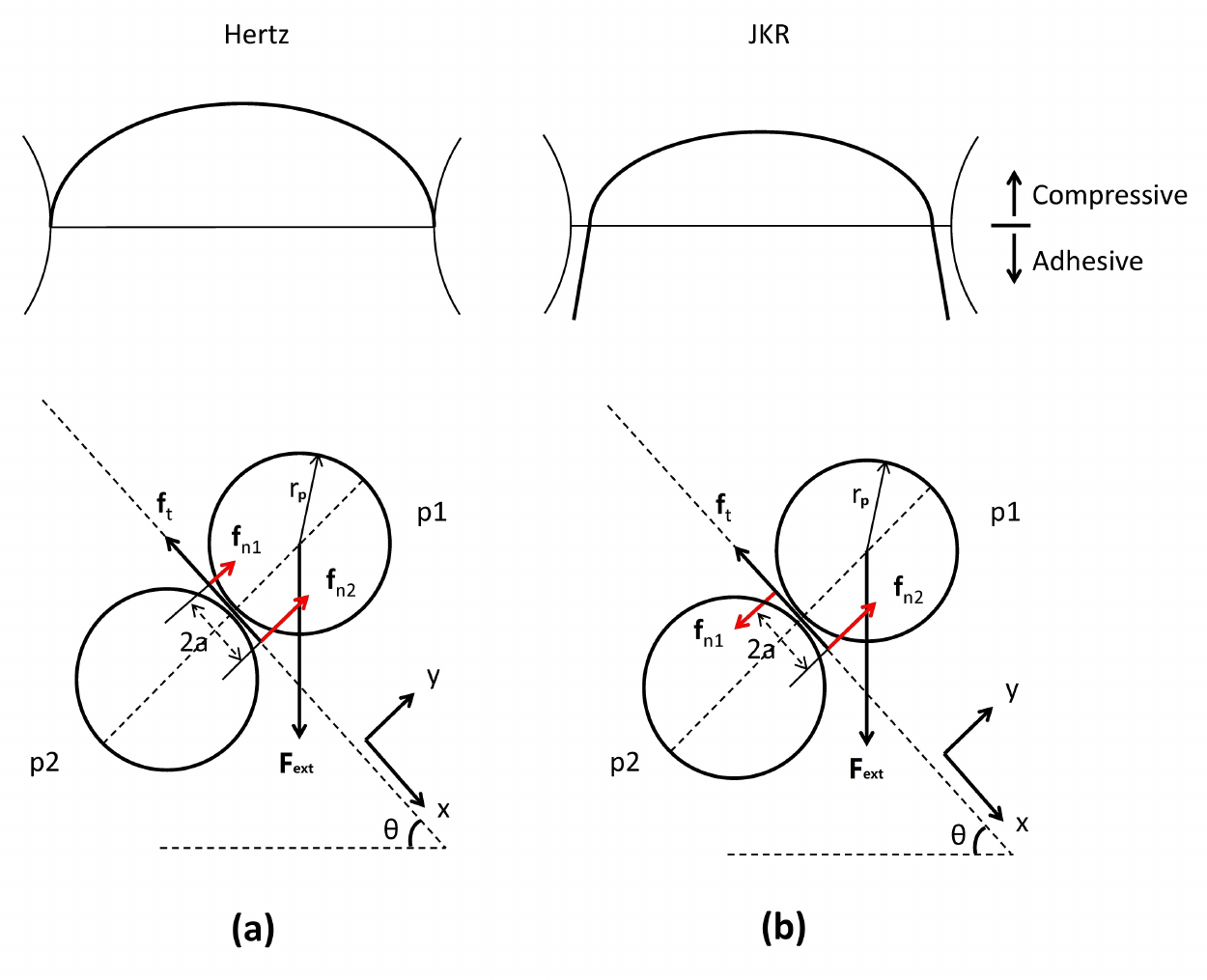}
\caption{\label{Fig_fschematic}(Colors online) Schematic diagrams illustrating the stress distributions and force balance of two contact particles for Hertz (a) and JKR (b) contact models. }
\end{center}
\end{figure}

\begin{figure}
\begin{center}
\includegraphics[width=7.5cm]{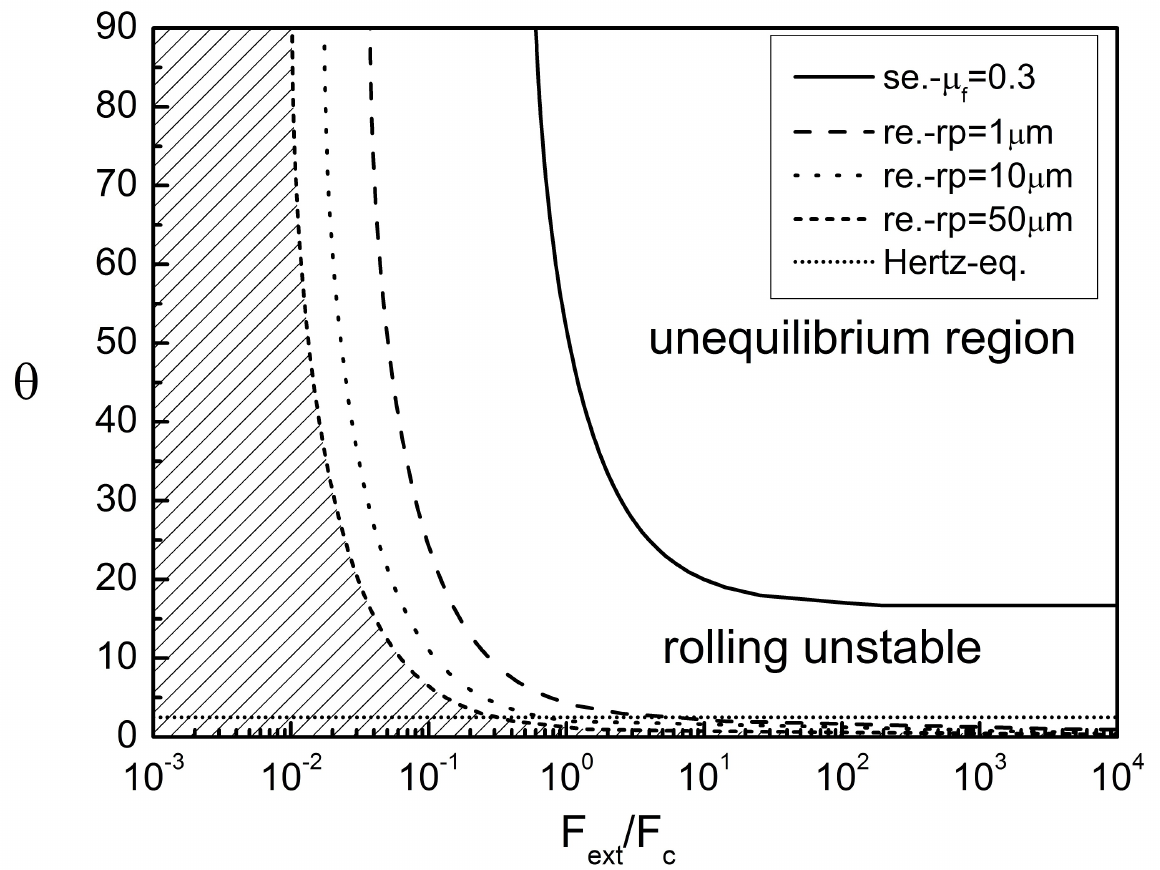}
\caption{\label{Fig_mecheql}(Colors online) Equilibrium diagram of two contact particles with JKR contact model. The solid line stands for the sliding equilibrium (se.) line. The dash, dot and short dash lines are the rolling equilibrium lines (re.) for the particles in our simulations with $r_p=1\mu m$, $r_p=10\mu m$ and $r_p=50\mu m$ respectively. The short dot line represents the equilibrium line of Hertz contact model. The shady area indicates the equilibrium region of $r_p=50\mu m$. }
\end{center}
\end{figure}

For convenience and better illustration, two contact particles $p1$ and $p2$ in 2D are drawn to interpret the mechanical equilibrium (see Fig.~\ref{Fig_fschematic}). $\theta$ is the angle between the contact surface and the horizontal and $a$ is the radius of the contact surface. Fig.~\ref{Fig_fschematic} also illustrates the stress distributions of Hertz and JKR contact models, which are used for adhesive-less granular matter and our adhesive packings, respectively. For both models, the stress distribution is symmetric. However, stress in Hertz model is totally repulsive while that in JKR model is repulsive in the central part and attractive on the rim of the contact surface. The unknown forces that act on $p1$ are the tangential force $f_t$ and two normal forces $f_{n1}$, $f_{n2}$ simplified by the stress distribution. When the angle $\theta$ and the external force $F_{ext}$ are known, two equations of force balance and one equation of torque balance can be solved to figure out the three unknown forces, indicating an isostatic condition. Here $F_{ext}$ could be gravity, electric force or other kinds of external forces that act on the particle. Usually, the particle $p1$ will start to slide or roll over $p2$ as $\theta$ and $F_{ext}$ reach some critical values, which is the cause of rearrangements in a packing. However, due to the presence of adhesive forces, the critical sliding force is greater than that of non-adhesive particles (see Eq.~\ref{fcrit}), meaning that particles with adhesion are more difficult to start sliding. For rolling, on the other hand, when two contact particles start to roll or have the tendency of rolling, the front side of the contact surface is compressed and the rear side will still be in touch until the critical pull-off force $F_C$ is reached, instead of detach immediately when there is no external force. As a consequence, the simplified normal forces $f_{n1}$, $f_{n2}$ will consist of a repulsive force on the front side of rolling and an attractive force on the rear side, while the two normal forces are both repulsive for adhesive-less particles (see the normal forces in Fig.~\ref{Fig_fschematic} indicated by the red arrows). The two normal forces of adhesive particles will generate torque of the same direction, resulting in greater rolling resistance. In this case, particles with adhesion can support larger external forces as well as be stabilized within a wider range of $\theta$ with constant external forces. Based on the above analysis, an equilibrium diagram is plotted with the parameters used in our work to demonstrate the equilibrium region in terms of angle $\theta$ and external force $F_{ext}$. Fig.~\ref{Fig_mecheql} shows two series of equilibrium lines, the rolling equilibrium (re.) line and the sliding equilibrium (se.) line, under which the area indicates the equilibrium region that the particle will not roll or slide over another. We can see that with very low external force, the adhesive particles can even be stabilized with $\theta=90^\circ$. Then when the external force grows larger, the angle $\theta$ decreases fast and reaches about $\theta \approx 0^\circ$ when the force further grows, implying that particles can only be stabilized when they are placed almost vertically. Also we found that the rolling equilibrium line lies under the sliding equilibrium line, shifting to left with increasing particle size, which means that rolling equilibrium breaks ahead of sliding and the larger size the particle is, the more easily it will roll. This is in agreement with the statement that sliding is relatively rare for small adhesive particles while rolling is generally the preferred deformation mode for agglomerates of adhesive particles \cite{Marshall14}. For comparison, the equilibrium line of Hertz model is also plotted in Fig.~\ref{Fig_mecheql} as indicated by the short dot line. The particle without adhesion can only reach equilibrium in a tiny range of $\theta(<2.5^\circ)$. This is because in Hertz model the normal force $f_{n1}$ on the rear side of the contact surface must be repulsive and thus cannot provide additional rolling resistance like that in JKR model, leading to probable rolling motion. After all, from the analysis above, we conclude that particles are able to get mechanical equilibrium with fewer neighbors than adhesive-less granular matter due to the existence of adhesion.

\section{Conclusions}

In this paper, random loose packings of uniform spherical micro-particles are investigated by using particle-level DEM simulations on the basis of adhesive contact mechanics. The packing structures arise from the competition between particle inertia and adhesion. A characteristic adhesion parameter ($Ad$) with respect to particle inertia is used to account for the combined effect of particle velocity, size and adhesion. When $1<Ad<20$, the volume fraction and coordination number are uniquely dependent on $Ad$, resulting in a new regime of adhesive loose packing. As $Ad$ grows above 20, the packing properties approach $\phi=0.125$, $Z=2$ and no longer changes with $Ad$, confirming the asymptotic adhesive loose packing limit. On the other hand, with $Ad<1$, indicating a weak adhesion, the packing properties go back to the range between RLP and RCP. A simple form of equation $\phi_{RCP}-\phi=\alpha Ad^\lambda$ is figured out to interpret the relationship between volume fraction and adhesion parameter, where the lowest volume fraction is achieved with $Ad>20$. Furthermore, the force distribution of adhesive packings resembles that of adhesive-less ones with $P(f)\sim f^\theta$ for small forces and $P(f)\sim \exp^{-\beta f}$ for big forces. The distinct exponents of $\theta=0.879$, $\beta=0.839$ found for normal force are supposed to be the results of the shrink of the force distribution. We conjecture that it is caused by the enhancement of very small forces and weakening of very big forces, respectively, in the presence of attractive adhesion forces. Mechanical equilibrium analysis shows that the attractive normal forces on the rear side of the contact surface of adhesive particles can provide additional rolling resistance such that they are more easily stabilized with lower coordination number than non-adhesive granular matter.

\section{Acknowledgements}
This work has been funded by the National Natural Science Funds of China (Nos. 50976058 and 51390491) and the National Key Basic Research Program of China (No. 2013CB228506). S. Q. Li is grateful to Prof. Jeff Marshall at University of Vermont, Prof. Hernan Makse at City College of New York and Dr. Guanqing Liu and Dr. Mengmeng Yang at Tsinghua University for helpful discussions. W. Liu acknowledges Shaojun Luo at City College of New York for his support of the simulation work.

\bibliography{refpre} 

\end{document}